\documentclass[11pt]{article}

\usepackage{color,graphicx}
\usepackage{young}
\usepackage[vcentermath]{youngtab}
\usepackage{amsmath,amssymb,graphicx}
\usepackage{hyperref}
\definecolor{darkred}{rgb}{0.65,0.15,0}
\hypersetup{pdfborder={0 0 0},colorlinks=true,urlcolor=darkred,citecolor=blue,linkcolor=darkred,linktocpage=true}

\usepackage{cite}
\usepackage{amsmath}
\usepackage{amsfonts}
\usepackage{amssymb}
\usepackage{graphicx}%
\usepackage{amsthm}
\usepackage{mathrsfs}
\usepackage[T1]{fontenc}
\usepackage{enumerate}
\usepackage{color}

\def\4diml{four-dimensional}

\def\-1{^{-1}}

\newcommand{\G}{\mathscr{G}}

\makeatletter

\@addtoreset{equation}{section}
\makeatother

\begin{document}

\thispagestyle{empty}

\vspace{5mm}

\begin{center}
{\LARGE \bf Yang-Baxter deformation of WZW model based on \\[2mm] Lie supergroups: The cases of $GL(1|1)$ and $(C^3 +A)$}

\vspace{15mm}
\normalsize
{\large  Ali Eghbali\footnote{Corresponding author: eghbali978@gmail.com}, Tayebe Parvizi\footnote{t.parvizi@azaruniv.ac.ir}, Adel Rezaei-Aghdam\footnote{rezaei-a@azaruniv.ac.ir}
}

\vspace{2mm}
{\small \em Department of Physics, Faculty of Basic Sciences,\\
Azarbaijan Shahid Madani University, 53714-161, Tabriz, Iran}\\

\vspace{7mm}


\vspace{6mm}

\begin{tabular}{p{12cm}}
{\small
We proceed to generalize the Yang-Baxter (YB) deformation of Wess-Zumino-Witten (WZW) model to the Lie supergroups case.
This generalization enables us to utilize various kinds of solutions of the
(modified) graded classical Yang-Baxter equation ((m)GCYBE) to classify
the YB deformations of WZW models based on the Lie supergroups.
We obtain the inequivalent solutions (classical r-matrices)
of the (m)GCYBE for the $gl(1|1)$ and $({\cal C}^3 +{\cal A})$ Lie superalgebras
in the non-standard basis,
in such a way that the corresponding automorphism transformations are employed.
Then, the YB deformations of the WZW models based on the $GL(1|1)$ and $(C^3 + A)$ Lie supergroups
are specified by skew-supersymmetric
classical r-matrices satisfying (m)GCYBE.
In some cases for both families of deformed models,
the metrics remain invariant under the deformation, while the components of $B$-fields are changed.
After checking the conformal invariance of the models up to one-loop order, it is concluded that
the $GL(1|1)$ and $(C^3 + A)$ WZW models are
conformal theories within the classes of the YB deformations preserving the conformal invariance.
However, our results are interesting in themselves,
but at a constructive level, may prompt many new insights into (generalized) supergravity solutions.
}
\end{tabular}
\vspace{-1mm}
\end{center}


~~~~~~~~~~~~~~~~~~~~~~Graded classical Yang-Baxter equation, Lie superalgebra

\setcounter{page}{1}
\newpage
\tableofcontents

 \vspace{5mm}

\section{\label{Sec.I} Introduction}

Klimcik \cite{{Klimcik1},{Klimcik2},{Klimcik3}} proposed the YB $\sigma$-model as a systematic way to consider integrable deformations of two-dimensional
non-linear $\sigma$-models.
Then, this systematic procedure refined by Delduc, Magro and Vicedo in \cite{Delduc1}.
The deformations obtained via this method are called {\it Yang-Baxter} deformations,
due to the central place that the CYBE takes in the construction.
The principal chiral models deformed by Klimcik were also generalized by Delduc, Magro and Vicedo in \cite{{Delduc2},{Delduc3}} for the
$AdS_5 \times S^5$ superstring action (see, also \cite{{Kawaguchi1},{Kawaguchi2},{Tongeren1}}).
In \cite{Delduc2}, the integrable deformation of the type $IIB$ $AdS_5 \times S^5$ superstring action
along with the deformed field equations, Lax connection, and $\kappa$-symmetry transformations have been presented.
Moreover, one can see the supercoset constructions in the YB deformed
$AdS_5 \times S^5$ superstring with the $SU(2,2|4)$ Lie supergroup based on the homogeneous CYBE in \cite{HidekiKyono} (see, also \cite{K.Yoshida}).
Actually, the integrable deformations of the $AdS_5 \times S^5$
superstring is an important application of the YB  $\sigma$-model description.
So far in all the works done on the deformation of the superstring action, the attention has been concentrated to the case where
the deformations are created by bosonic generators of the Lie supergroup.
Unlike these works, in the present work, the deformation is performed on both bosonic and fermionic sectors of the models.

The YB $\sigma$-model was been then generalized by adding a WZW term.
A prescription of the YB deformation of WZW model invented by Delduc, Magro and Vicedo in \cite{Delduc4} (see, also
\cite{Klimcik2017,Quantum,2020}).
In most of the works, the deformations of the YB WZW models have been studied on semisimple or compact Lie groups.
Some interesting examples of the deformed YB WZW models were constructed on
the Lie groups Nappi-Witten \cite{kyono2016yang}, $H_{4}$ \cite{Epr} and $GL(2,\mathbb{R})$ \cite{Epr2}
with classical r-matrices satisfying the (m)CYBE.
A fundamental fact about them is that all can be considered as unique conformal theories within the class of the YB
deformations preserving the conformal invariance.

The goal of the present work is to generalize the YB deformation of WZW model to the Lie supergroups case
and present the resulting YB deformed backgrounds for the $GL(1|1)$ and $(C^3 +A)$ Lie supergroups along with inequivalent
classical r-matrices satisfying the (m)GCYBE.
This generalization would be important from the viewpoint of its applications, because the YB deformed backgrounds on the Lie supergroups
have a wider class of the (generalized) supergravity solutions \cite{Tsytlin} in general rather than the bosonic Lie groups.

This paper is organized as follows. In Section 2, by introducing a useful notation of ${\mathbb{Z}}_{2}$-graded vector space
we generalize the YB deformation of WZW model to the Lie supergroups case.
In Section 3, we find the $R$-operators and inequivalent r-matrices for the $gl(1|1)$ Lie superalgebra.
We furthermore construct the YB deformed backgrounds of
the $GL(1|1)$ WZW model in this section. Calculating inequivalent r-matrices for the $({\cal C}^3 +{\cal A})$ Lie superalgebra
and followed by the YB deformations of the $(C^3 +A)$ WZW model are devoted to Section 4.
In Section 5, it is shown that the deformed backgrounds satisfy the one-loop beta function equations which is the most
important feature of the obtained models. In this way, we obtain the dilaton fields making the deformed models conformal up to the one-loop order.
Some concluding remarks are given in the last section.

\section{\label{Sec.II} YB deformation of WZW model based on Lie supergroups and (m)GCYBE}

We are now interested in studying the YB deformation of WZW model based on Lie supergroups.
The general procedure that we shall apply is a straightforward generalization of
the well-known prescription of Delduc, Magro and Vicedo \cite{Delduc4}.
Thus, in this section, inspired by a prescription invented by authors of Ref. \cite{Delduc4}, we
generalize the YB deformation of WZW model from Lie groups to Lie supergroups.
Before setting the model with Lie supergroups,  let us recall the properties of $\mathbb{Z}_2$-graded vector
space and also the definition of a Lie superalgebra $\G$ \cite{N.A}.
A super vector space $V$ is a ${\mathbb{Z}}_{2}$-graded vector space, i.e., a vector space over a field
$\mathbb{K}$ with a given decomposition of subspaces of grade 0 and grade 1, $V= V_{_0} \oplus V_{_1}$.
The parity of a nonzero homogeneous element, denoted by $|x|$, is $0$ (even) or $1$ (odd)\footnote{The even elements are sometimes called bosonic, and the odd elements fermionic. From now on,
we use $B$ and $F$ instead of $0$ and $1$, respectively.} according to whether it is in $V_{0}$ or $V_{1}$, namely,
$|x|=0$ for any $x \in V_{_0}$, while for any $x \in V_{_1}$ we have $|x|=1$.
A Lie superalgebra ${\G}$ is a $\mathbb{Z}_2$-graded vector space, thus admitting the decomposition
${\G} ={\G}_{{_B}} \oplus {\G}_{_F}$, equipped with a bilinear
superbracket structure $[. , .]: {\cal G} \otimes {\cal G} \rightarrow {\cal G}$ satisfying the requirements of anti-supersymmetry
and super Jacobi identity.
If ${\G}$ is finite-dimensional and the dimensions of ${\G}_{_B}$ and $ {\G}_{_F}$
are $m=\#B$ and  $n=\#F$, respectively, then ${\G}$ is said to have dimension $(m|n)$.
We shall identify grading indices by the same indices in the power of $(-1)$, i.e., we use $(-1)^x$ instead of  $(-1)^{|x|}$, where
$(-1)^x$ equals 1 or -1 if the Lie sub-superalgebra element is even or odd, respectively\footnote{
Note that this notation was first used by Dewitt in \cite{D}. Throughout this paper we work with Dewitt's notation}.

Let us turn our attention to the model setting with Lie supergroups. First of all, it should be noted that
the original WZW model based on a Lie supergroup $G$ in Dewitt's notation was first presented in \cite{ER7}. Accordingly,
the action of the YB deformed  WZW model on a Lie supergroup $G$
may be expressed as\footnote{The last term in
\eqref{2.1} is the standard WZ term integrated over a 3-dimensional space $B_{3}$  parameterized
by $(\tau, \sigma, \xi)$ and whose boundary is the worldsheet ${\Sigma}$, where
the extra direction is labeled by $\xi$. In this term, $\varepsilon_{_{\alpha\beta\gamma}}$  is the Levi-Civita symbol in  three dimensions.}
\begin{eqnarray}\label{2.1}			
S^{^{YB }}_{_{WZW}}(g)=\frac{1}{2}\int_{_\Sigma}d\sigma^+ d\sigma^-  ~ (-1)^a J_{+}^{a} \Omega_{_{ab}} L_{-}^{b}  +\frac{\kappa}{12}\
\int_{_{B_{3}}}d^{3}\sigma  \varepsilon^{\alpha\beta\gamma} ~ (-1)^{a+bc}  L_{\alpha}^{a} \Omega_{ad}~ f^d_{~bc} ~ L_{\beta}^{b} L_{\gamma}^{c},~~
\end{eqnarray}
where $\sigma^\alpha = (\sigma^+ , \sigma^-)$ are the standard lightcone variables such that their relationship with
the worldsheet coordinates $(\tau , \sigma)$ is given by $\sigma^{\pm} = {(\tau\pm\sigma)}/\sqrt{2}$.
Here, the left-invariant super one-form $L_{\alpha} = g^{-1} ~\partial_{\alpha} g$
is written in terms of an element $g(\tau, \sigma)$ of the Lie supergroup $G$. $L_{\alpha}$ is a ${\G}$-valued function, that is, it
can be written as $L_{\alpha} =(-1)^a  L_{\alpha}^{a} T_{_a}$, in which $T_{_a}, a=1,...,dim \hspace{0.4mm}G$ are the basis of Lie superalgebra ${\G}$ of $G$.
A key ingredient contained in both terms of the action \eqref{2.1} is the most general
non-degenerate invariant supersymmetric bilinear form $\Omega_{_{ab}}$ on the Lie algebra ${\G}$ which satisfies
the following condition \cite{ER7}:
\begin{eqnarray}\label{2.2}
f^d_{~ab} \;\Omega_{dc}+(-1)^{bc} f^d_{~ac} \;\Omega_{db}\;=\;0.
\end{eqnarray}
Note that the bilinear form $\Omega_{_{ab}}$ is defined as inner product $<. ~,~.>$ for the basis $T_{_a}$ of $\G$, and
$f^c_{~ab}$ are the structure constants which determine the (anti-)commutation relations $[T_{_a} , T_{_b}]=f^c_{~ab}~ T_{_c}$. The deformed currents $J_{\pm} =(-1)^a  J_{\pm}^{a} T_{_a} $ are defined in the following form
\begin{eqnarray}\label{2.3}
J_{\pm} = (1+\omega \eta^{2})\frac{1 \pm \tilde{A} R}{1-\eta^{2}R^{2}} L_{\pm},
\end{eqnarray}
where $\eta, \tilde A$ and  $\kappa$ are three independent real parameters such that the
deformation is measured by $\eta$ and $\tilde A$. The last parameter $\kappa$ is regarded as the level.
When $\eta= \tilde A=0$ and $\kappa=1$, the action \eqref{2.1} is nothing but that of the original WZW model on the Lie supergroup \cite{ER7}.
The operator $R$ in \eqref{2.3} is a linear map
from the Lie superalgebra $\G$ to itself, $R: {\G} \rightarrow {\G}$. It is a
skew-supersymmetric solution of the (m)GCYBE on ${\G}$. That is to say, for any $X, Y \in \G$ it satisfies
\begin{eqnarray}\label{2.4}
[R(X),R(Y)]-R\big([R(X),Y]+[X,R(Y)]\big)=\omega [X,Y].
\end{eqnarray}
Here $\omega$ is a constant parameter which can be normalized by rescaling $R$. Equation \eqref{2.4} can be generalized to
the mGCYBE if one sets $\omega=\pm1$, while the case with $\omega=0 $ is the homogeneous GCYBE. Moreover,
the skew-supersymmetric condition of the linear $R$-operator requires that
\begin{eqnarray}\label{2.5}
<R(X) , Y>+ <X  , R(Y)> =0.
\end{eqnarray}
In what follows we will focus on a class of linear $R$-operators constructed from a
classical $r$-matrix $r \in {\cal G} \otimes {\cal G}$ by means of the general formula\footnote{We note
that the inner product is evaluated on the second site of the
r-matrix.}
\begin{eqnarray}\label{2.6}
R(X)=<r ~, ~1 \otimes X>,
\end{eqnarray}
for any $X \in \G$. Here the r-matrix defined as  $r = r^{ab}~ T_{_a} \otimes T_{_b}$ is a solution of the following standard (m)GCYBE \cite{N.A}
\begin{eqnarray}\label{2.7.1}
[[r , r]]\equiv [r_{_{12}} , r_{_{13}}] +[r_{_{12}} , r_{_{23}}]+[r_{_{13}} , r_{_{23}}] =\omega ~\Omega,
\end{eqnarray}
where $r_{_{12}} =r \otimes 1$, $r_{_{23}} =1 \otimes r$ and $r_{_{13}} = r^{ab}~ T_{_a} \otimes 1 \otimes T_{_b}$;
moreover, $\Omega \in \Lambda^3 (\G)$ is the canonical triple tensor Casimir of G.
Notice that the standard form of the (m)GCYBE is equivalent to \eqref{2.4}.
When the $r$-matrix is a skew-supersymmetric solution of \eqref{2.7.1}, i.e., $r^{ab} =- (-1)^{ab}~ r^{ba}$, one can write
\begin{eqnarray}\label{2.7}
r&=&\frac{1}{2}  r^{ab} \big(T_{_a} \otimes T_{_b} -(-1)^{ab}~ T_{_b} \otimes T_{_a}\big)\nonumber\\
&=& \frac{1}{2} r^{ab} ~T_{_a} \wedge T_{_b}.
\end{eqnarray}
We furthermore note that the $r$-matrix is considered to be even as $r \in \G_{_B} \wedge \G_{_B} \oplus \G_{_F} \wedge  \G_{_F}$
so that it has the following matrix representation\footnote{For a Lie superalgebra $\G=\G_B \oplus  \G_F$ of dimension $(m|n)$ we define
the basis of $\G$ as $\{T_{_a}\}_{a=1}^{m+n}=\{t_{_i}, S_{_\alpha}\}$ where $\{t_{_i}\}_{i=1}^{m}$ and
$\{S_{_\alpha}\}_{\alpha=m+1}^{m+n}$ are the bosonic and fermionic basis, respectively.
Accordingly, the r-matrix can be written into the form
\begin{eqnarray}\label{2.8.1}
r= {r}_{_B}^{ij}  ~ t_i \otimes t_j + {r}_{_F}^{\alpha \beta} ~ S_\alpha  \otimes S_\beta.
\end{eqnarray}
}
\begin{eqnarray}\label{2.8}
r^{ab}=\left( \begin{tabular}{c|c}
                 $r_{_B}$ & 0 \\\hline
                 0 & $r_{_F}$ \\
                 \end{tabular} \right).
\end{eqnarray}
According to this, $r^{ab} =0$ if $|a| \neq |b|$.
In other words, fermions with bosons can't be mixed (grading is preserved).
By using the fact that in $r^{ab}$, $|a|+|b|=0$, and
by expanding $X$ and $R$ in terms of the bases of $\G$ as $X= (-1)^{a}~ x^{a} T_{_a}$ and $R= (-1)^{b}~ R_a^{~b} T_{_b}$, and then by substituting
\eqref{2.7} in \eqref{2.6} we find
\begin{eqnarray}\label{2.9}
R_a^{~b}=-(-1)^{ac} ~\Omega_{ac}  ~  r^{cb}.
\end{eqnarray}
Matrices such as $\Omega_{ab}$ and $R_a^{~b}$ are also considered similar to \eqref{2.8}, that is,
one considers for them $|a|+|b|=0$.
Accordingly, the (m)GCYBE \eqref{2.4} can be rewritten into the following form:
\begin{eqnarray}\label{2.10}
(-1)^k~R_a^{~c}~f^k_{~cd} R_b^{~d}-(-1)^b~R_a^{~c}~f^d_{~cb} R_d^{~k}-(-1)^a~R_b^{~c}~f^d_{~ac} R_d^{~k} = \omega  ~ f^k_{~ab}.
\end{eqnarray}
It is also useful to obtain matrix form of the above equation by using the matrix representations of the structure constants,
$f^c_{~ab} =- ({\cal Y}^c)_{ ab}$, giving us\footnote{Here the superscript ``st'' in $R^{^{st}}$ stands for supertranspose \cite{D}.}
\begin{eqnarray}\label{2.11}
(-1)^d~R~{\cal Y}^k R^{^{st}}-(-1)^c~R ({\cal Y}^d R_d^{~k})- ({\cal Y}^d R_d^{~k}) R^{^{st}} =(-1)^k~ \omega {\cal Y}^k,
\end{eqnarray}
where index $d$ in the first term of the left hand side denotes the column of matrix ${\cal Y}^k$,
while in the second sentence, $c$ corresponds to the row  of
matrix ${\cal Y}^d$.
In the next sections, we employ the above formulation in order to obtain the linear $R$-operators and r-matrices on
the $gl(1|1)$ and $({\cal C}^3 +{\cal A})$ Lie superalgebras.
By using the obtained $R$-operators we will find all YB deformations of WZW models based on these Lie supergroups.


\section{\label{Sec.III} YB deformations of WZW model on the $GL(1|1)$ Lie supergroup}

In this section we first solve the (m)GCYBE \eqref{2.11} in order to obtain the
$R$-operators and inequivalent r-matrices for the $gl(1|1)$ Lie superalgebra. Using
the resulting $R$-operators we construct the YB deformed backgrounds of
the $GL(1|1)$ WZW model.

\subsection{\label{Sec.III.1} $R$-operators and r-matrices of the $gl(1|1)$}

First of all, let us introduce the $gl(1|1)$ Lie superalgebra.
In Backhouse's classification \cite{B}, the $gl(1|1)$  has been denoted by $({\cal C}_{-1}^2 + {\cal A})$; in fact,
traditional notation for the $(C^2_{-1}+A)$ Lie superalgebra is
the $gl(1|1)$. On the other, in Ref. \cite{ER4} we classified all four-dimensional
Drinfeld superdoubles of the type $(2 | 2)$ and showed that
there are just three classes of non-isomorphic Drinfeld superdoubles of the type $(2 | 2)$ so that two of them are isomorphic to the
Lie superalgebras $gl(1|1)$ and $({\cal C}^3 + {\cal A})$, another is an Abelian Lie superalgebra.
These possess two bosonic generators and two fermionic ones.
We shall denote the former by $T_{_1}, T_{_2}$ and use $T_{_3}, T_{_4}$ for fermionic generators.
From now on we consider $T_{_1}, T_{_2}$ and $T_{_3}, T_{_4}$ as bosonic and fermionic generators, respectively.
For the $gl(1|1)$, the relations between these elements are, in the non-standard basis, given by \cite{B}
\begin{equation}\label{3.1}
[T_{_1} , T_{_3}] = T_{_3},\;\;\;[T_{_1} , T_{_4}] = -T_{_4},\;\;\{T_{_3} , T_{_4}\}
=T_{_2}.
\end{equation}
It should be noted that in Ref. \cite{ER6} two of us obtained all
Lie superbialgebra structures on the $gl(1|1)$ and their corresponding r-matrices in the standard basis.
According to DeWitt's notation \cite{D}, in the standard basis the structure constants $f^B_{_{FF}}$ are considered to be
pure imaginary.
As we showed a moment ago in \eqref{3.1} here we work in the non-standard basis,
so our results on the Lie superbialgebra structures and corresponding r-matrices will be different from those of \cite{ER6}.
The $gl(1|1)$ Lie superalgebra possesses a non-degenerate supersymmetric ad-invariant metric $\Omega_{ab}$
which is defined for any pair of bases $T_{_a}, T_{_b} \in gl(1|1)$ such that by using \eqref{2.2} and also the structure constants of \eqref{3.1}
one gets \cite{ER7}
\begin{eqnarray}\label{3.2}
\Omega_{ab}=\left( \begin{tabular}{cccc}
                $\beta$ & $\alpha$  &  0        & 0\\
                $\alpha$     & 0  &  0        & 0\\
                0     & 0  &  0        & $\alpha$\\
                0     & 0  & $-\alpha$ & 0\\
                 \end{tabular} \right),
\end{eqnarray}
for some real constants $\alpha, \beta$.
The metric is needed e.g. to write down the action of WZW model on the $GL(1|1)$ Lie supergroup.

Before proceeding to solve the (m)GCYBE \eqref{2.11} for the $gl(1|1)$, let us first assume that
the most general skew-supersymmetric r-matrix $r \in {\cal G}_{_{(2|2)}} \otimes {\cal G}_{_{(2|2)}}$ has the following form:
\begin{eqnarray}\label{3.3.1}
r=m_1 T_{_1} \wedge T_{_2} +m_2 T_{_3} \wedge T_{_4} +\frac{1}{2} m_3 T_{_3} \wedge T_{_3}+\frac{1}{2} m_4 T_{_4} \wedge T_{_4}.
\end{eqnarray}
Comparing this with \eqref{2.7} one can obtain the matrix representation of
$r^{ab}$, giving us
\begin{eqnarray}\label{3.3}
r^{ab}=\left( \begin{tabular}{cccc}
                0 & $m_1$  &  0  & 0\\
                $-m_1$     & 0  &  0        & 0\\
                0     & 0  &  $m_3$  & $m_2$\\
                0     & 0  & $m_2$ & $m_4$\\
                 \end{tabular} \right),
\end{eqnarray}
where $m_i$ are some real parameters. In addition,
the matrix representations of the $gl(1|1)$ are easily obtained to be
{{\small \begin{eqnarray}\label{3.4}
{\cal Y}^1 = 0,~~~{\cal Y}^2 =\left(\begin{array}{cccc}
0 & 0  & 0 &  0\\
0 & 0  & 0 &  0\\
0 & 0  & 0 & -1\\
0 & 0  & -1 & 0
\end{array}
\right),~~~{\cal Y}^3 =\left(\begin{array}{cccc}
0 & 0  & -1 &  0\\
0 & 0  & 0 &  0\\
1 & 0  & 0 & 0\\
0 & 0  & 0 & 0
\end{array}
\right),~~~{\cal Y}^4 = \left(\begin{array}{cccc}
0 & 0  & 0 &  1\\
0 & 0  & 0 &  0\\
0 & 0  & 0 & 0\\
-1 & 0  & 0 & 0
\end{array}
\right).
\end{eqnarray}}}
Inserting \eqref{3.2} and \eqref{3.3} into \eqref{2.9} one can obtain the general form of the corresponding $R$-operator. Thus,
by substituting the resulting $R$-operator and also the representations \eqref{3.4} into equation \eqref{2.11},
the general solution of the (m)GCYBE is split into four families ${R_{_I}}_{_a}^{~b}$, ${R_{_{II}}}_{_a}^{~b}$,
${R_{_{III}}}_{_a}^{~b}$ and ${R_{_{IV}}}_{_a}^{~b}$
such that the solutions are, in terms of the constants
$\alpha, \beta, \omega$,  $m_i$, given by
{\small \begin{eqnarray}\nonumber
{R_{_I}}_{_a}^{~b}=\left( \begin{tabular}{cccc}
                $\alpha m_1$ & $-\beta m_1$  &  0  & 0\\
                0     & $-\alpha m_1$  &  0        & 0\\
                0     & 0  & $\pm \sqrt{-\omega}$  & 0\\
                0     & 0  & 0 & $\mp \sqrt{-\omega}$\\
                 \end{tabular} \right),~~~~{R_{_{II}}}_{_a}^{~b}=\left( \begin{tabular}{cccc}
                0 & 0  &  0  & 0\\
                0    & 0  &  0 & 0\\
                0     & 0  &  0  & $-\frac{\omega}{\alpha m_3}$\\
                0     & 0  & $-\alpha m_3$ & 0\\
                 \end{tabular} \right),
\end{eqnarray}}
\vspace{-2mm}
{\small \begin{eqnarray}\label{3.5}
{R_{_{III}}}_{_a}^{~b}=\pm \sqrt{-\omega} \left( \begin{tabular}{cccc}
                1 & -$\frac{\beta}{\alpha}$  &  0  & 0\\
                0     & -1  &  0        & 0\\
                0     & 0  &  1  & $ \pm \frac{\alpha m_4}{\sqrt{-\omega}}$\\
                0     & 0  & 0 & -1\\
                 \end{tabular} \right),~~~{R_{_{IV}}}_{_a}^{~b}=\pm \sqrt{-\omega} \left( \begin{tabular}{cccc}
                -1 & $\frac{\beta}{\alpha}$  &  0  & 0\\
                0     & 1  &  0        & 0\\
                0     & 0  &  1  & 0\\
                0     & 0  & $\mp \frac{\alpha m_3}{\sqrt{-\omega}}$ & -1\\
                 \end{tabular} \right).~~
\end{eqnarray}}
Again by employing \eqref{2.9} and \eqref{3.2} one can obtain the corresponding r-matrices in the form of \eqref{2.8.1}, giving us
\begin{eqnarray}\label{3.6}
{r_{_I}}&=& m_1  ~ T_{_1} \wedge T_{_2}  \pm \frac{\sqrt{-\omega}}{\alpha} ~ T_{_3}  \wedge T_{_4},~~~~~~~~~~~~~~~~~~~\nonumber\\
{r_{_{II}}}&=&\frac{m_3}{2}  ~ T_{_3} \wedge T_{_3}  - \frac{\omega}{2 \alpha^2 m_3} ~ T_{_4}  \wedge T_{_4},\nonumber\\~~~~~~~~~~~~~~~~~~~
{r_{_{III}}}&=&\pm \frac{\sqrt{-\omega}}{\alpha} \Big(T_{_1} \wedge T_{_2}+T_{_3} \wedge T_{_4}\Big)  + \frac{m_4}{2} ~ T_{_4}  \wedge T_{_4},\nonumber\\~~~~~~~~~~~~~~~~~~~
{r_{_{IV}}}&=&\mp \frac{\sqrt{-\omega}}{\alpha} \Big(T_{_1} \wedge T_{_2}-T_{_3} \wedge T_{_4}\Big)  + \frac{m_3}{2} ~ T_{_3}  \wedge T_{_3}.~~~~~~~~~~~~~~~~~~~
\end{eqnarray}
The next step is that to determine the inequivalent r-matrices for the $gl(1|1)$.
In fact, we need to specify the exact value of the parameters $m_i$ of the solutions \eqref{3.6}.
In Ref. \cite{Epr} as a Proposition we proved that
two r-matrices $r$ and $r'$ of a Lie algebra $\G$ are equivalent if one can be obtained from the other by means of
a change of basis which is an automorphism $A$ of Lie algebra $\G$. Here we generalize the Proposition to the super case.\\\\
{\bf Proposition 3.1.} {\it Two r-matrices $r$ and $r'$ of a Lie superalgebra $\G$ are equivalent
if there exists an automorphism $A$ of $\G$ such that
\begin{eqnarray}\label{3.7}
r^{ab} = (-1)^d~ (A^{^{st}})^a_{~c}~ {r^\prime}^{cd}~{A_d}^b.
\end{eqnarray} }
The proof of this Proposition is similar to those of \cite{Epr}.

According to formula \eqref{3.7} in order to obtain the inequivalent r-matrices one must use the automorphism group of Lie superalgebra $\G$
which preserves (a) the parity of the generators (they can't mix fermions with bosons), and (b) the structure constants $f^c_{~ab}$.
Therefore it is crucial for our further considerations to identify the
supergroup of automorphisms of the $gl(1|1)$ Lie superalgebra.
We define the action of the automorphism $A$ on $\G$ by the transformation $T'_a = (-1)^b~ {A}_a^{~b}~ T_b$.
The set of automorphisms of $gl(1|1)$ is generated by two transformations:
\begin{eqnarray}
T'_1 = -T_1+c T_2,~~~~~T'_2 = ab T_2,~~~~~~T'_3 = -a T_4,~~~~~~T'_4 = -b T_3,~~~~\label{3.8}
\end{eqnarray}
and
\begin{eqnarray}
T'_1 = T_1+c T_2,,~~~~~~T'_2 = ab T_2,~~~~~~T'_3 = -a T_3,~~~~~~T'_4 = -b T_4,~~~~\label{3.9}
\end{eqnarray}
where $a, b, c$ are arbitrary real numbers such that $ab\neq 0$.
The bases $\{T'_a\}$ obey the same (anti-)commutation relations as
$\{T_a\}$. When taken into account, the above transformations lead to a conclusion that the parameters $m_i$ in \eqref{3.6}
can be scaled out to take the value
of 1 or 0. Now by using the automorphism transformations and by employing formula \eqref{3.7},
one can determine the inequivalent r-matrices for the $gl(1|1)$.
Finally we arrive at eleven families of inequivalent r-matrices whose representatives can be described
by means of the following Theorem.\\\\
{\bf Theorem 3.1.}~{\it Any r-matrix of the $gl(1|1)$ Lie superalgebra as a solution of the (m)GCYBE belongs
just to one of the following eleven inequivalent classes\footnote{
As we mentioned at the beginning of subsection \eqref{Sec.III.1},  all
Lie superbialgebra structures on the $gl(1|1)$ and their corresponding r-matrices have been, in the standard basis, obtained  in \cite{ER6}.
There, it has been shown that among seventeen families of inequivalent Lie superbialgebra
structures, only six of them are of coboundary type, while in the present work we have obtained
eleven families of inequivalent r-matrices. The reason behind this is that if one solves
super co-Jacobi and mixed super Jacobi identities for the $gl(1|1)$ in the non-standard basis, then he/she sees that the
solutions will be different from those of \cite{ER6}.
}}
\begin{eqnarray*}
{r_{_i}}&=& T_{_1} \wedge T_{_2},~~~~~~~~~~~~~~~~~~~\nonumber\\
{r_{_{ii}}}&=&\frac{1}{2} T_{_3}  \wedge T_{_3},\nonumber\\~~~~~~~~~~~~~~~~~~~
{r_{_{iii}}}&=&-\frac{1}{2} T_{_3}  \wedge T_{_3},\nonumber\\~~~~~~~~~~~~~~~~~~~
{r_{_{iv}}}&=& T_{_3}  \wedge T_{_4},\nonumber\\~~~~~~~~~~~~~~~~~~~
{r_{_{v}}}&=& T_{_1} \wedge T_{_2} + m_2  T_{_3}  \wedge T_{_4},~~~m_2=\frac{\sqrt{-\omega}}{\alpha}>0,~~~m_2 \neq 1,\nonumber\\
{r_{_{vi}}}&=& \frac{1}{2} \big(T_{_3} \wedge T_{_3} +  T_{_4}  \wedge T_{_4}\big),\nonumber\\
{r_{_{vii}}}&=& \frac{1}{2} \big(T_{_3} \wedge T_{_3} -  T_{_4}  \wedge T_{_4}\big),\nonumber\\
{r_{_{viii}}}&=& -\frac{1}{2} \big(T_{_3} \wedge T_{_3} +  T_{_4}  \wedge T_{_4}\big),\nonumber\\
{r_{_{ix}}}&=& T_{_1} \wedge T_{_2} +  T_{_3}  \wedge T_{_4},\nonumber\\
{r_{_{x}}}&=& T_{_1} \wedge T_{_2} +T_{_3} \wedge T_{_4} + \frac{1}{2} T_{_4}  \wedge T_{_4},\nonumber\\
{r_{_{xi}}}&=& T_{_1} \wedge T_{_2} +T_{_3} \wedge T_{_4} - \frac{1}{2} T_{_4}  \wedge T_{_4}.\nonumber
\end{eqnarray*}
It should be noted that among eleven inequivalent classes of the r-matrices, only ${r_{_i}}$, ${r_{_{ii}}}$ and ${r_{_{iii}}}$ satisfy the
standard GCYBE with $\omega =0$, while the rest are solutions of the mGCYBE with $\omega =-\alpha^2$ except for
${r_{_{v}}}$ and ${r_{_{vii}}}$ which is $\omega =\alpha^2$ for ${r_{_{vii}}}$.
The parameter $m_2$ is present in ${r_{_{v}}}$ as it can't be removed by means of the transformations \eqref{3.8} and \eqref{3.9}.
It means that for different values of $m_2$ we have inequivalent r-matrices. However, as we will see,
the $m_2$ plays a role of the deformation parameter in the YB deformed background of the $GL(1|1)$ WZW model.

Before closing this subsection, let us look at the unimodularity condition on the solutions of the (m)GCYBE for the $gl(1|1)$,
Theorem 3.1. As we know the r-matrix is the initial input for construction of the YB deformed backgrounds.
When the r-matrix satisfies the unimodularity condition that is given by \cite{Wulff}
\begin{eqnarray}
r^{ab}~ [T_a , T_b] =0,\label{3.10}
\end{eqnarray}
then, the resulting deformed background is a solution to type IIB supergravity. If not, the background does not satisfy
the on-shell condition of the supergravity and becomes a solution to a generalized supergravity.
Below we determine which of the r-matrices classified in Theorem 3.1 are unimodular and or non-unimodular.
Using \eqref{3.10} together with \eqref{3.1} we find that the r-matrices ${r_{_{iv}}}, {r_{_{v}}}, {r_{_{ix}}}, {r_{_{x}}}$
and ${r_{_{xi}}}$ are non-unimodular, while the rest denote the unimodular r-matrices.
In the following, by calculating the linear $R$-operators corresponding to the inequivalent r-matrices of the $gl(1|1)$
we will deformed the $GL(1|1)$ WZW model.


\subsection{\label{Sec.III.2} YB deformed backgrounds of the $GL(1|1)$ WZW model}

Before proceeding to construct out the YB deformed backgrounds of
the $GL(1|1)$ WZW model, let us have an overview of undeformed WZW model structure
based on the $GL(1|1)$ Lie supergroup.
In Ref. \cite{ER7}, it was constructed the $GL(1|1)$ WZW model in order to study super Poisson-Lie symmetry \cite{ER2} of the model.
As mentioned in section \ref{Sec.II}, by setting $\eta= \tilde A=0$ and $\kappa=1$ in \eqref{2.3} one gets the original WZW model from
the action \eqref{2.1}.
Let us introduce a supergroup element represented by
\begin{eqnarray}
g = e^{\chi T_4}~e^{y T_1}~e^{x T_2}~e^{\psi T_3}, \label{3.11}
\end{eqnarray}
where $x(\tau , \sigma)$ and $y(\tau , \sigma)$ denote bosonic fields
while $\psi(\tau , \sigma)$ and $\chi(\tau , \sigma)$ stand for fermionic fields.
Using \eqref{3.11}, the components of left-invariant super one-form $L^{a}_{\pm}$ on the $GL(1|1)$ can be evaluated as \cite{ER7}
\begin{eqnarray}
L^{1}_{\pm} &=& \partial_{\pm} y,~~~~~~~~~~~~~~~~~~~~~~~ L^{2}_{\pm} = \partial_{\pm} x -  \partial_{\pm} \chi ~\psi e^y,\nonumber\\
L^{3}_{\pm} &=&- \partial_{\pm} \psi -  \partial_{\pm}y ~\psi,~~~~~~~~ L^{4}_{\pm} = -\partial_{\pm} \chi~e^y.\label{3.12}
\end{eqnarray}
A key ingredient in writing down the action of a WZW model is the most general supersymmetric ad-invariant form such that for
the $gl(1|1)$ has been given by equation \eqref{3.2}. Finally by using \eqref{3.1}, \eqref{3.2} and  \eqref{3.12}
one can write down the action of WZW model based on the $GL(1|1)$ similar to what was done in Ref. \cite{ER7}.
The corresponding supersymmetric metric and anti-supersymmetric two-form field ($B$-field) are given by
\begin{eqnarray}\label{3.13}
ds^2 &=& (-1)^{\mu \nu} ~G_{_{\mu \nu}} dx^\mu~dx^\nu = \beta dy^2+ 2dydx -2 e^y~ d\psi d\chi,\nonumber\\
B &=& \frac{1}{2} (-1)^{\mu \nu} ~B_{_{\mu \nu}} dx^\mu  \wedge dx^\nu =-e^y~ d\psi \wedge d\chi.
\end{eqnarray}
Here we have assumed that the constant $\alpha$ of $\Omega_{ab}$ in \eqref{3.2} is set to be $1$.
From now on we consider $\alpha=1$.
Equation \eqref{3.13} as a background of the WZW model should be conformally invariant.
To check this, one first looks at the one-loop beta function equations \cite{ER5}
\begin{eqnarray}
&&{\cal R}_{_{\mu\nu}}+\frac{1}{4}H_{_{\mu\rho\sigma}} {H^{^{\sigma \rho}}}_{_{\nu}}+2{\overrightarrow{\nabla}_{_\mu}}
{\overrightarrow{\nabla}_{_\nu}} \Phi ~=~0,\nonumber\\
&&(-1)^{^{\lambda}} {\nabla}^{^\lambda}\big(e^{^{-2 \Phi}} H_{_{\lambda\mu\nu}}\big)
~=~0,\nonumber\\
&&4 \Lambda-{\cal R}-\frac{1}{12} H_{_{\mu\nu\rho}}H^{^{\rho\nu\mu}} +4{\overrightarrow{\nabla}_{_\mu}} {\Phi} \overrightarrow{\nabla}^{^\mu} {\Phi}
 - 4 {\overrightarrow{\nabla}_{_\mu}} \overrightarrow{\nabla}^{^\mu} {\Phi}  =0,\label{c.5}
\end{eqnarray}
where the covariant derivatives ${\overrightarrow {\nabla}}_{_\mu}$, scalar curvature ${\cal R}$ and Ricci tensor ${\cal R}_{_{\mu\nu}}$
are calculated from the metric $G_{_{\mu\nu}}$ that is also used for lowering and raising
indices, and $H_{_{\mu\nu\rho}}$ is the field strength corresponding to the $B$-field which is defined by
\begin{eqnarray}\label{a.1.1}
H_{_{\mu\nu\rho}} = (-1)^{\mu} \frac{\overrightarrow {\partial}}{\partial x^{\mu}} B_{_{\nu \rho}} +(-1)^{\nu +\mu ( \nu+ \rho)} \frac{\overrightarrow {\partial}}{\partial x^{\nu}} B_{_{\rho \mu}}+(-1)^{\rho +\rho( \mu+ \nu)} \frac{\overrightarrow {\partial}}{\partial x^{\rho}} B_{_{\mu\nu}}.
\end{eqnarray}
For the background \eqref{3.13} one easily verifies equations \eqref{c.5} with a constant dilaton field, ${\Phi} = \varphi_0$, and vanishing cosmological constant.

Let us turn into the  main goal of this subsection which is nothing but calculating the
YB deformations of the $GL(1|1)$ WZW model.
As we mentioned earlier, having $R$-operators one can calculate the deformed currents.
Now we use formulas \eqref{2.9} and \eqref{3.2}
to obtain all linear $R$-operators corresponding to the inequivalent r-matrices of Theorem 3.1.
In order to calculate the currents $J_{\pm}$ one may write down relation \eqref{2.3} in the following form
\begin{eqnarray}\label{3.29}
J^a_{\pm} -(-1)^{b+c}~\eta^{2} J^b_{\pm} ~R_b^{~c}~R_c^{~a}= (1+\omega \eta^{2})\big[L^{a}_{_\pm} \pm (-1)^{b}~ \tilde{A} L^{b}_{_\pm} ~R_b^{~a}\big].
\end{eqnarray}


\begin{center}
		\small {{{\bf Table 1.}~ YB deformed backgrounds of the $GL(1|1)$ WZW model}}
		{\scriptsize
			\renewcommand{\arraystretch}{1.5}{
\begin{tabular}{|p{1.85cm}|l|l|} \hline \hline
Background symbol & ~~~~Backgrounds including metric and $B$-field & Comments\\ \hline

{$GL(1|1)_{i}^{(\eta,\tilde{A},\kappa)}$}~~~ &  ~~~~$ds^{2}=\frac{1}{1-\eta^{2}}\Big[\beta dy^{2}+2 dy dx	
+{2\eta^{2}}\psi e^{y}dy d\chi\Big]-2 e^{y} d\psi d\chi$ & \\&  ~~~~$B=\frac{\tilde{A}}{1-\eta^{2}}~\psi e^{y}dy\wedge d\chi-  \kappa e^{y} d\psi\wedge d\chi$ &$\omega=0$\\\hline

{$GL(1|1)_{ii}^{(\tilde{A}, \kappa)}$} &  ~~~~$ds^{2}=\beta dy^{2}+2 dy dx-2 e^{y} d\psi d\chi $ & \\&
~~~~$B=-\frac{1}{2} \tilde{A} e^{2y}~ d\chi\wedge d\chi - \kappa e^{y} d\psi\wedge d\chi$ & $\omega=0$ \\\hline
	
{$GL(1|1)_{iii}^{(\tilde{A}, \kappa)}$} & ~~~~$ds^{2}=\beta dy^{2}+2dy dx-2 e^{y} d\psi d\chi$ &\\& ~~~~
$B=\frac{1}{2} \tilde{A} e^{2y}~ d\chi\wedge d\chi - \kappa e^{y} d\psi\wedge d\chi$  &  $\omega=0$\\  \hline

{$GL(1|1)_{iv}^{(\eta, \kappa)}$} & ~~~~$ds^{2}=(1-\eta^{2})\Big(\beta dy^{2}+2dy dx\Big)-2\eta^{2} \psi e^{y}dy d\chi  -2 e^{y} d\psi d\chi$ &\\& ~~~~
$B=- \kappa e^{y} d\psi\wedge d\chi$  &  $\omega=-1$\\\hline

$GL(1|1)_{v}^{(\eta, \tilde{A}, \kappa)}$ & ~~~~$ds^{2}= \frac{1}{1-\eta^{2}}\Big[\beta (1-m_{2}^{2}\eta^{2}) dy^{2}+2 dy dx+2\eta^{2}(1-m_{2}^{2})\psi e^{y}dy d\chi\Big] -2 e^{y} d\psi d\chi$ &\\& ~~~~
$B=\frac{\tilde{A}(1-m_{2}^{2}\eta^{2})}{1-\eta^{2}}~\psi e^{y}dy\wedge d\chi-\kappa e^{y} d\psi\wedge d\chi$  &  $m_2=\sqrt{-\omega}$\\\hline
	
$GL(1|1)_{vi}^{(\eta, \tilde{A}, \kappa)}$ & ~~~~$ds^{2}= (1-\eta^{2})\Big[\beta dy^{2}+2 dy dx
+\frac{2 \eta^{2}}{1+\eta^{2}}\psi e^{y}dy d\chi-\frac{2}{1+\eta^{2}} e^{y} d\psi d\chi\Big]$ &\\& ~~~~
$B=\frac{-\tilde{A}(1-\eta^{2})}{1+\eta^{2}}\big[\psi dy\wedge d\psi+\frac{1}{2}e^{2y} d\chi \wedge d\chi\big]- \kappa e^{y} d\psi\wedge d\chi$  &  $\omega=-1$\\\hline

$GL(1|1)_{vii}^{(\eta, \tilde{A}, \kappa)}$ & ~~~~$ds^{2}= (1+\eta^{2})\Big[\beta dy^{2}+2 dy dx -\frac{2 \eta^{2}}{1-\eta^{2}}\psi e^{y}dy d\chi-\frac{2}{1-\eta^{2}} e^{y} d\psi d\chi\Big]$ &\\& ~~~~
$B=\frac{\tilde{A}(1+\eta^{2})}{1-\eta^{2}}~[\psi dy\wedge d\psi-\frac{1}{2}e^{2y} d\chi \wedge d\chi ]-\kappa e^{y} d\psi\wedge d\chi$  &  $\omega=1$\\\hline

$GL(1|1)_{viii}^{(\eta, \tilde{A}, \kappa)}$ & ~~~~$ds^{2}=(1-\eta^{2})\Big[\beta dy^{2}+2 dy dx +\frac{2\eta^{2}}{1+\eta^{2}}\psi e^{y}dy d\chi-\frac{2}{1+\eta^{2}} e^{y} d\psi d\chi\Big]$ &\\& ~~~~
$B=\frac{\tilde{A}(1-\eta^{2})}{1+\eta^{2}}~[\psi dy\wedge d\psi+\frac{1}{2}e^{2y}d\chi \wedge d\chi ]-\kappa e^{y} d\psi\wedge d\chi$  &  $\omega=-1$\\\hline

$GL(1|1)_{ix}^{(\tilde{A},\kappa)}$ & ~~~~$ds^{2}= \beta dy^{2}+2 dy dx-2 e^{y} d\psi d\chi$ &\\& ~~~~
$B=\tilde{A}~\psi e^{y}dy\wedge d\chi- \kappa e^{y} d\psi\wedge d\chi$  &  $\omega=-1$\\\hline

$GL(1|1)_{x}^{(\tilde{A},\kappa)}$ & ~~~~$ds^{2}= \beta dy^{2}+2 dy dx-2 e^{y} d\psi d\chi$ &\\& ~~~~
$B=-\tilde{A}~\psi dy\wedge d\psi -  (\kappa+ \tilde{A} ) e^{y} d\psi\wedge d\chi$  &  $\omega=-1$\\\hline

$GL(1|1)_{xi}^{(\tilde{A},\kappa)}$ & ~~~~$ds^{2}=\beta dy^{2}+2 dy dx-2e^{y} d\psi d\chi$ &\\& ~~~~
$B=\tilde{A}~\psi dy\wedge d\psi - (\kappa+\tilde{A} ) e^{y} d\psi\wedge d\chi$  &  $\omega=-1$\\\hline\hline
		\end{tabular}}}
\end{center}
Finally by using the resulting linear $R$-operators satisfying the (m)GCYBE, and also by utilizing relation
\eqref{3.29} together with \eqref{3.12} one obtains the YB deformations of the $GL(1|1)$ WZW model.
The deformed backgrounds including metric and $B$-field together with the related comments
are summarized in Table 1.
Notice that the symbol of each background, e.g. $GL(1|1)_{i}^{(\eta,\tilde{A},\kappa)}$,
indicates the deformed background derived by $r_{_{{i}}}$; roman numbers $i$, $ii$ etc.
distinguish between several possible deformed backgrounds of the $GL(1|1)$ WZW model,
and the $(\kappa, \eta, \tilde{A})$ indicate the deformation parameters of each background.

As it is seen from Table 1, in some of the backgrounds such as $GL(1|1)_{ii}^{(\tilde{A},\kappa)}$,
$GL(1|1)_{iii}^{(\tilde{A},\kappa)}$,
$GL(1|1)_{ix}^{(\tilde{A},\kappa)}$,  $GL(1|1)_{x}^{(\tilde{A},\kappa)}$
and $GL(1|1)_{xi}^{(\tilde{A},\kappa)}$, the metrics are invariant under the deformation, up to two-form $B$-fields.
That is, the effect coming from the deformations is reflected only as the coefficient of B-field.
With the exceptions of the $GL(1|1)_{ii}^{(\tilde{A},\kappa)}$, $GL(1|1)_{iii}^{(\tilde{A},\kappa)}$ and $GL(1|1)_{iv}^{(\eta,\kappa)}$,
for the rest of the backgrounds
we have ignored the total derivative terms that appeared in the $B$-fields part.

\section{\label{Sec.IV} YB deformations of WZW model on the $(C^3 +A)$ Lie supergroup}

Similarly to the performance of calculations for the $gl(1|1)$,
in this section we first solve the (m)GCYBE \eqref{2.11} to obtain the
$R$-operators and inequivalent r-matrices for the  $({\cal C}^3 +{\cal A})$ Lie superalgebra.
We then get YB deformations of the WZW model based on
the $(C^3 +A)$ Lie supergroup by utilizing the inequivalent r-matrices satisfying the (m)GCYBE.
This is the subject of the present section.

\subsection{\label{Sec.IV.1} $R$-operators and r-matrices of the $({\cal C}^3 +{\cal A})$}

The $({\cal C}^3 +{\cal A})$ Lie superalgebra is spanned by the set of generators $\{T_{_1}, T_{_2}; T_{_3}, T_{_4}\}$ which
fulfill the following  non-zero (anti-)commutation rules \cite{B}:
\begin{equation}\label{4.1}
[T_{_1} , T_{_4}] = T_{_3},\;\;\;\;\{T_{_4} , T_{_4}\}
=T_{_2}.
\end{equation}
Notice that the Lie superbialgebra structures on the $({\cal C}^3 +{\cal A})$ along with their corresponding r-matrices, in the standard basis,
were obtained in \cite{ER14}. Here we work in the non-standard basis; accordingly, our results on
the r-matrices will be different from those of \cite{ER14}.

Analogously, we consider an element $r \in ({\cal C}^3 +{\cal A}) \otimes ({\cal C}^3 +{\cal A})$ as in \eqref{3.3.1}, or equivalently, \eqref{3.3}.
On the other hand, using \eqref{2.2} one easily checks that the non-degenerate ad-invariant metric on the $({\cal C}^3 +{\cal A})$ is the same
\eqref{3.2}. The general form of the corresponding $R$-operator can be found by inserting \eqref{3.2} and \eqref{3.3} into \eqref{2.9}.
Calculating the matrix representations $({\cal Y}^c)_{ ab}$ of the $({\cal C}^3 +{\cal A})$
and then putting the resulting $R$-operator
into \eqref{2.11}, the most general solution can be determined like
\begin{eqnarray}\label{4.2}
{R}_{_a}^{~b}=\left( \begin{tabular}{cccc}
                $m_1$ & $-\beta m_1$  &  0  & 0\\
                0     & $-m_1$  &  0        & 0\\
                0     & 0  & $m_2$  & 0\\
                0     & 0  & $-m_3$ & $-m_2$\\
                \end{tabular} \right).
\end{eqnarray}
Here the condition \eqref{2.11} has led to the following constraints:
\begin{eqnarray}\label{4.3}
\omega=m_2 (m_2+2 m_1),~~~m_4=0.
\end{eqnarray}
Again by employing \eqref{2.9}, the corresponding r-matrix to the above solution is obtained to be
\begin{eqnarray}\label{4.4}
{r}= m_1 T_{_1} \wedge T_{_2}+ m_2 T_{_3} \wedge T_{_4}  + \frac{1}{2} m_3 ~ T_{_3}  \wedge T_{_3}.
\end{eqnarray}
In the following, in order to find inequivalent r-matrices we need to specify the exact value of the parameters $m_i$ of the above solution.
For this purpose, one must use the formula \eqref{3.7}. The use of this formula requires that we know the automorphism
transformation of the given Lie superalgebra.
For the $({\cal C}^3 +{\cal A})$ the automorphism transformation preserving the (anti)commutation rules \eqref{4.1} is given by
\begin{eqnarray}\label{4.5}
T'_1 = a T_1+c T_2,~~~~~T'_2 = b^2 T_2,~~~~~~T'_3 = -ab T_3,~~~~~~T'_4 = -d T_3- bT_4,
\end{eqnarray}
for some constants $a, b, c, d$. After performing the transformation \eqref{4.5} on formula \eqref{3.7},
one concludes that r-matrices of the $({\cal C}^3 +{\cal A})$ are split into eight inequivalent classes.
For the sake of clarity the results are summarized in Theorem 4.1.
\\\\
{\bf Theorem 4.1.}~{\it Any r-matrix of the $({\cal C}^3 +{\cal A})$ Lie superalgebra as a solution of the (m)GCYBE belongs
just to one of the following eight inequivalent classes}
\begin{eqnarray*}
{r_{_{i}}}&=&\frac{1}{2} T_{_3}  \wedge T_{_3},\nonumber\\~~~~~~~~~~~~~~~~~~~
{r_{_{ii}}}&=&-\frac{1}{2} T_{_3}  \wedge T_{_3},\nonumber\\~~~~~~~~~~~~~~~~~~~
{r_{_{iii}}}&=& T_{_3} \wedge T_{_4},~~~~~~~~~~~~~~~~~~~\nonumber\\
{r_{_{iv}}}&=& T_{_1} \wedge T_{_2},\nonumber\\~~~~~~~~~~~~~~~~~~~
{r_{_{v}}}&=& T_{_1} \wedge T_{_2} + \frac{1}{2} T_{_3}  \wedge T_{_3},\nonumber\\
{r_{_{vi}}}&=&  T_{_1} \wedge T_{_2} - \frac{1}{2} T_{_3}  \wedge T_{_3},\nonumber\\
{r_{_{vii}}}&=& T_{_1} \wedge T_{_2} + m_2~ T_{_3}  \wedge T_{_4},~~~\omega =m_2(m_2 +2),~~m_2 \neq 0, -2\nonumber\\
{r_{_{viii}}}&=& T_{_1} \wedge T_{_2} -2~ T_{_3}  \wedge T_{_4}.\nonumber
\end{eqnarray*}
It is noteworthy that only the r-matrices ${r_{_{iii}}}$ and ${r_{_{vii}}}$  satisfy the
mGCYBE with $\omega =1$ and $\omega =m_2(m_2 +2)$, respectively, while the rest are solutions of the GCYBE.
At the end of this subsection it should be noted that all inequivalent r-matrices above
are unimodular, that is, they satisfy the unimodularity condition \eqref{3.10}.

\subsection{\label{Sec.IV.2} YB deformed backgrounds of the $(C^3 +A)$ WZW model}
We start this subsection by introducing the $(C^3 +A)$ WZW model.
The $(C^3 +A)$ WZW model based on the $(C^3 +A)$ Lie supergroup was originally created in Ref. \cite{ER8}
in order to study its super Poisson-Lie T-dualizability \cite{ER2}.
In order to write the model explicitly we need to find the super one-form $L^{a}_{\pm}$'s. To this purpose we
use a general element of $(C^3 +A)$ as in \eqref{3.11}. Then we find \cite{ER8}
\begin{eqnarray}
L^{1}_{\pm} &=& \partial_{\pm} y,~~~~~~~~~~~~~~~~~~~~~~~ L^{2}_{\pm} = \partial_{\pm} x -  \partial_{\pm} \chi ~\frac{\chi}{2},\nonumber\\
L^{3}_{\pm} &=&- \partial_{\pm} \psi +  \partial_{\pm} \chi~y,~~~~~~~~ L^{4}_{\pm} = -\partial_{\pm} \chi.\label{4.6}
\end{eqnarray}
As mentioned before, one must set the parameters
$\eta= \tilde A=0$ and $\kappa=1$ in \eqref{2.3} to get the original WZW model from
the action \eqref{2.1}. Using \eqref{4.1}, \eqref{4.6} and the fact that the ad-invariant metric on the $({\cal C}^3 +{\cal A})$ is the same
\eqref{3.2}, one computes the action of WZW model on the $(C^3 +A)$ Lie supergroup.
From the action one can easily read off the corresponding metric and anti-supersymmetric fields, giving us
\begin{eqnarray}\label{4.7}
ds^2 &=&  \beta dy^2+ 2dydx + \chi dy d\chi-2  d\psi d\chi,\nonumber\\
B &=&  \frac{\chi}{2}~ dy \wedge d\chi.
\end{eqnarray}
Indeed, this background satisfies the one-loop beta function equations \eqref{c.5} with ${\Phi} = \varphi_0$ and $\Lambda=0$.

\begin{center}
		\small {{{\bf Table 2.}~ YB deformed backgrounds of the $(C^3 +A)$ WZW model}}
		{\scriptsize
			\renewcommand{\arraystretch}{1.5}{
\begin{tabular}{|p{1.92cm}|l|l|} \hline \hline
Background symbol & ~~~~Backgrounds including metric and $B$-field & Comments\\ \hline

{${(C^3 +A)}_{i}^{(\tilde{A},\kappa)}$}~~~ &  ~~~~$ds^{2}=\beta dy^2+ 2dydx + \chi dy d\chi-2  d\psi d\chi$ & \\&
~~~~$B=-\frac{1}{2} \tilde{A}~  d\chi \wedge d\chi + \frac{1}{2} \kappa \chi~ dy \wedge d\chi$ &$\omega=0$\\\hline

{${(C^3 +A)}_{ii}^{(\tilde{A},\kappa)}$}~~~ &  ~~~~$ds^{2}=\beta dy^2+ 2dydx + \chi dy d\chi-2  d\psi d\chi$ & \\&
~~~~$B=+\frac{1}{2} \tilde{A}~  d\chi \wedge d\chi +\frac{1}{2} \kappa \chi~ dy \wedge d\chi$ &$\omega=0$\\\hline
	
{${(C^3 +A)}_{iii}^{(\eta, \tilde{A}, \kappa)}$} & ~~~~$ds^{2}=(1+\eta^{2}) \Big[\beta dy^2+
 2dydx + \chi dy d\chi-\frac{2}{(1-\eta^{2})} d\psi d\chi \Big]$ &\\& ~~~~
$B=-\frac{\tilde{A} (1+\eta^{2})}{(1-\eta^{2})} y ~ d\chi \wedge d\chi +\frac{1}{2} \kappa \chi~ dy \wedge d\chi$  &  $\omega=1$\\  \hline

{${(C^3 +A)}_{iv}^{(\eta, \tilde{A}, \kappa)}$} & ~~~~
$ds^{2}=\frac{1}{(1-\eta^{2})} \Big[\beta dy^{2}+2dy dx+ \chi dy d\chi\Big]-2 d\psi d\chi$ &\\& ~~~~
$B=\frac{1}{2} \Big[\kappa + \frac{\tilde{A}}{(1-\eta^{2})}\Big] \chi  dy \wedge d\chi$  &  $\omega=0$\\\hline

{${(C^3 +A)}_{v}^{(\eta, \tilde{A}, \kappa)}$} & ~~~~
$ds^{2}=\frac{1}{(1-\eta^{2})} \Big[\beta dy^{2}+2dy dx+ \chi dy d\chi\Big]-2 d\psi d\chi$ &\\& ~~~~
$B=\frac{1}{2} \Big[\kappa + \frac{\tilde{A}}{(1-\eta^{2})}\Big] \chi  dy \wedge d\chi -\frac{1}{2} \tilde{A}~  d\chi \wedge d\chi$  &  $\omega=0$\\\hline
	
{${(C^3 +A)}_{vi}^{(\eta, \tilde{A}, \kappa)}$} & ~~~~
$ds^{2}=\frac{1}{(1-\eta^{2})} \Big[\beta dy^{2}+2dy dx+ \chi dy d\chi\Big]-2 d\psi d\chi$ &\\& ~~~~
$B=\frac{1}{2} \Big[\kappa + \frac{\tilde{A}}{(1-\eta^{2})}\Big] \chi  dy \wedge d\chi+\frac{1}{2} \tilde{A}~  d\chi \wedge d\chi$  &  $\omega=0$\\\hline

${(C^3 +A)}_{vii}^{(\eta, \tilde{A}, \kappa)}$ & ~~~~$ds^{2}=\frac{(1+ \omega \eta^{2})}{(1-\eta^{2})}\Big[\beta dy^{2}+2 dy dx +\chi~dy d\chi\Big]
-\frac{2(1+ \omega \eta^{2})}{(1-m_{2}^{2} \eta^{2})}  d\psi d\chi$ & $\omega =m_2(m_2 +2)$\\& ~~~~
$B=\frac{1}{2} \Big[\kappa + \frac{\tilde{A} (1+ \omega \eta^{2})}{(1-\eta^{2})}\Big] \chi  dy \wedge d\chi
-\frac{ \tilde{A} m_{2} (1+ \omega \eta^{2})}{(1-m_{2}^{2} \eta^{2})} y  d\chi \wedge d\chi$
&  $~~m_2 \neq 0, -2$\\\hline

${(C^3 +A)}_{viii}^{(\eta, \tilde{A}, \kappa)}$ & ~~~~$ds^{2}=\frac{1}{(1-\eta^{2})}\Big[\beta dy^{2}+2 dy dx +\chi~dy d\chi\Big]
-\frac{2}{1-4 \eta^{2}}  d\psi d\chi$ &\\& ~~~~
$B=\frac{1}{2} \Big[\kappa + \frac{\tilde{A}}{(1-\eta^{2})}\Big] \chi  dy \wedge d\chi
+ \frac{2 \tilde{A}}{(1-4 \eta^{2})} y  d\chi \wedge d\chi $  &  $\omega=0$\\\hline\hline
		\end{tabular}}}
\end{center}

We are looking for our main goal in this section, which is nothing but calculating the
YB deformations of the $(C^3 +A)$ WZW model.
First, employing formulas \eqref{2.9} and \eqref{3.2}
we obtain all linear $R$-operators corresponding to the inequivalent r-matrices of Theorem 4.1.
Then, making use of the relations \eqref{3.29} and \eqref{4.6} one obtains the deformed currents $J_{\pm}$.
Finally we have used the action \eqref{2.1} to classify all YB deformed backgrounds of the $(C^3 +A)$ WZW model.
The results including metric and $B$-field are summarized in Table 2.
As it is seen, only in the backgrounds ${(C^3 +A)}_{i}^{(\tilde{A},\kappa)}$ and ${(C^3 +A)}_{ii}^{(\tilde{A},\kappa)}$,
the metrics remained unchanged under transformation, up to the $B$-fields.
In addition, for all backgrounds we have ignored the total derivative terms that appeared in the $B$-fields part, except for
the mentioned backgrounds.


\section{\label{Sec.V} Conformal invariance of the YB deformed backgrounds}

Our goal in this section is to investigate the conformal invariance conditions of the deformed models. In fact,
we shall show that the WZW models based on the $GL(1|1)$ and $(C^3 + A)$ Lie supergroups can be considered as
conformal theories within the classes of the YB deformations preserving the conformal
invariance up to the one-loop order. Accordingly, using the equations \eqref{c.5} we check the conformal invariance conditions of
the deformed backgrounds (Tables 1 and 2).
From solving the equations we find the general form
of the dilaton fields that make the deformed backgrounds conformal up to the one-loop order.
The results obtained for the deformations of the $GL(1|1)$ WZW model are represented in Table 3.
It is noteworthy that in all cases the cosmological constant vanishes.
Also, the results obtained from solving equations \eqref{c.5} for the deformed backgrounds of the $(C^3 + A)$ WZW model
are summarized in Table 4.
In some cases of the $(C^3 +A)$ deformed backgrounds, we have shown that dilton fields can depend on both bosonic coordinates.
Note that $c_0$ and $c_1$ in Tables 3 and 4 are some arbitrary constants.
\begin{center}
		\scriptsize {{{\bf Table 3.}~  The dilaton fields making the $GL(1|1)$ deformed backgrounds conformal up to one-loop order}}
		{\scriptsize
			\renewcommand{\arraystretch}{1.5}{
\begin{tabular}{|p{1.85cm}|l|l|} \hline \hline
Background symbol & ~~~~Dilaton field & Comments\\ \hline

{$GL(1|1)_{i}^{(\eta,\tilde{A},\kappa)}$}~~~ &  ~~~~$\Phi=\frac{\Gamma}{8(1-\eta^{2})^{2}}y^{2}+c_{_1}y+c_{_0}$ & $\Gamma= \Big[\tilde{A}+\kappa (1-\eta^{2})\Big]^{2}-1$ \\\hline

{$GL(1|1)_{ii}^{(\tilde{A},\kappa)}$}~~~ &  ~~~~$\Phi=\frac{1}{8}(\kappa^{2}-1)y^{2}+c_{_1}y+c_{_0}$ &  \\\hline

{$GL(1|1)_{iii}^{(\tilde{A},\kappa)}$}~~~ &  ~~~~$\Phi=\frac{1}{8}(\kappa^{2}-1)y^{2}+c_{_1}y+c_{_0}$ &  \\\hline

{$GL(1|1)_{iv}^{(\eta,\kappa)}$}~~~ &  ~~~~$\Phi=\frac{1}{8}\big[\kappa^2 -(1-\eta^{2})^{2}\big] y^{2}+c_{_1}y+c_{_0}$ &  \\\hline

{$GL(1|1)_{v}^{(\eta,\tilde{A},\kappa)}$}~~~ &  ~~~~$\Phi=\frac{\Gamma}{8(1-\eta^{2})^{2}}y^{2}+c_{_1}y+c_{_0}$ & $\Gamma= \Big[\tilde{A} (1-m_{2}^{2}\eta^{2})+\kappa (1-\eta^{2})\Big]^{2}-(1-m_{2}^{2}\eta^{2})^2$ \\\hline

{$GL(1|1)_{vi}^{(\eta,\tilde{A},\kappa)}$}~~~ &  ~~~~$\Phi=\frac{\Gamma}{2}y^{2}+c_{_1}y+c_{_0}$ & $\Gamma= \tilde{A}^{2}-\frac{(1+\eta^{2})^{2}}{4(1- \eta^{2})^{2}}\big[(1-\eta^{2})^{2}-\kappa^{2}\big]$ \\\hline

{$GL(1|1)_{vii}^{(\eta,\tilde{A},\kappa)}$}~~~ &  ~~~~$\Phi=\frac{\Gamma}{2}y^{2}+c_{_1}y+c_{_0}$ & $\Gamma= -\tilde{A}^{2}-\frac{(1-\eta^{2})^{2}}{4(1+ \eta^{2})^{2}}\big[(1+\eta^{2})^{2}-\kappa^{2}\big]$ \\\hline

{$GL(1|1)_{viii}^{(\eta,\tilde{A},\kappa)}$}~~~ &  ~~~~$\Phi=\frac{\Gamma}{2}y^{2}+c_{_1}y+c_{_0}$ & $\Gamma= \tilde{A}^{2}-\frac{(1+\eta^{2})^{2}}{4(1- \eta^{2})^{2}}\big[(1-\eta^{2})^{2}-\kappa^{2}\big]$ \\\hline

{$GL(1|1)_{ix}^{(\tilde{A},\kappa)}$}~~~ &  ~~~~$\Phi=\frac{1}{8}\big[(\tilde{A}+\kappa)^{2}-1\big] y^{2}+c_{_1}y+c_{_0}$ &  \\\hline

{$GL(1|1)_{x}^{(\tilde{A}, \kappa)}$} &  ~~~~$\Phi=\frac{1}{8}\big[(\tilde{A}+\kappa)^{2}-1\big] y^{2}+c_{_1}y+c_{_0}$ &  \\\hline

{$GL(1|1)_{xi}^{(\tilde{A}, \kappa)}$} &  ~~~~$\Phi=\frac{1}{8}\big[(\tilde{A}+\kappa)^{2}-1\big] y^{2}+c_{_1}y+c_{_0}$ &  \\\hline

		\end{tabular}}}
\end{center}


\begin{center}
		\scriptsize {{{\bf Table 4.}~ The dilaton fields making the $(C^3 +A)$ deformed backgrounds conformal up to one-loop order}}
		{\scriptsize
			\renewcommand{\arraystretch}{1.5}{
\begin{tabular}{|p{1.92cm}|l|l|} \hline \hline
Background symbol & ~~~~Dilaton field & Comments\\ \hline

{${(C^3 +A)}_{i}^{(\tilde{A},\kappa)}$}~~~ &  ~~~~$\Phi=c_{_1}y+c_{_0}$ & $\Lambda=0,$ \\\hline

{${(C^3 +A)}_{ii}^{(\tilde{A},\kappa)}$}~~~ &  ~~~~$\Phi=c_{_1}y+c_{_0}$ & $\Lambda=0,$ \\\hline
	
{${(C^3 +A)}_{iii}^{(\eta, \tilde{A}, \kappa)}$} & ~~~~$\Phi=c_{_1}y+c_{_0};$ & $\Lambda=0,$ \\& ~~~
$\Phi= \Big(\frac{\Lambda (1+\eta^{2})}{\beta}\Big)^{\frac{1}{2}} x+c_{_0};$  &  $\tilde{A} =-\frac{\kappa (1-\eta^{2})}{2 (1+\eta^{2})}$\\  \hline

{${(C^3 +A)}_{iv}^{(\eta, \tilde{A}, \kappa)}$} & ~~~~
$\Phi=c_{_1}y+c_{_0};$ & $\Lambda=0,$ \\& ~~~
$\Phi= \Big(\frac{\Lambda}{\beta (1-\eta^{2})}\Big)^{\frac{1}{2}} x+c_{_0};$  &  $\tilde{A} =-{5(1-\eta^{2})}$\\  \hline

{${(C^3 +A)}_{v}^{(\eta, \tilde{A}, \kappa)}$} & ~~~~
$\Phi=c_{_1}y+c_{_0};$ & $\Lambda=0,$ \\& ~~~
$\Phi= \Big(\frac{\Lambda}{\beta (1-\eta^{2})}\Big)^{\frac{1}{2}} x+c_{_0};$  &  $\tilde{A} =-{5(1-\eta^{2})}$\\  \hline

{${(C^3 +A)}_{vi}^{(\eta, \tilde{A}, \kappa)}$} & ~~~~
$\Phi=c_{_1}y+c_{_0};$ & $\Lambda=0,$ \\& ~~~
$\Phi= \Big(\frac{\Lambda}{\beta (1-\eta^{2})}\Big)^{\frac{1}{2}} x+c_{_0};$  &  $\tilde{A} =-{5(1-\eta^{2})}$\\  \hline

${(C^3 +A)}_{vii}^{(\eta, \tilde{A}, \kappa)}$ & ~~~~$\Phi=c_{_1}y+c_{_0};$ & $\Lambda=0,$ \\& ~~~
$\Phi= \Big(\frac{\Lambda \big[1+m_2 (m_2 +2)\eta^{2}\big]}{\beta (1-\eta^{2})}\Big)^{\frac{1}{2}} x+c_{_0};$  &
$\tilde{A} =-\frac{\kappa (1-\eta^{2}) (1-m_{2}^{2} \eta^{2})}{\big[1+m_{2} (m_2+2) \eta^{2}\big]\big[1+2m_2-m_2(m_2+2)\eta^{2}\big]}$\\  \hline

${(C^3 +A)}_{viii}^{(\eta, \tilde{A}, \kappa)}$ & ~~~~$\Phi=c_{_1}y+c_{_0};$ & $\Lambda=0,$ \\& ~~~
$\Phi= \Big(\frac{\Lambda}{\beta (1-\eta^{2})}\Big)^{\frac{1}{2}} x+c_{_0};$  &  $\tilde{A} =\frac{\kappa}{3} (1-\eta^{2}) (1-4 \eta^{2})$\\  \hline\hline
		\end{tabular}}}
\end{center}
\section{Summary and concluding remarks}
\label{Sec.VI}
We have generalized the formulation of YB deformation of WZW model proposed by
Delduc, Magro and Vicedo from Lie groups to Lie supergroups.
As showed, this generalization enabled us to find the various kinds of the solutions to the (m)GCYBE.
As two influential examples, we classified the inequivalent r-matrices as solutions
of the (m)GCYBE for the $gl(1|1)$ and $({\cal C}^3 +{\cal A})$ Lie superalgebras
in the non-standard basis.
Using these solutions we could construct YB deformations of the WZW models based on the
$GL(1|1)$ and $({C}^3 +{A})$ Lie supergroups.
We furthermore showed that the deformed backgrounds
are conformally invariant up to the one-loop order which is the most
important feature of the resulting models.
With this interpretation, we have shown that the WZW models on the aforementioned supergroups can be considered as
conformal theories within the classes of the YB deformations preserving the conformal
invariance up to one-loop order.

As mentioned earlier, here we have worked with two of the WZW models based on the $GL(1|1)$ and $({C}^3 +{A})$ Lie supergroups.
The $GL(1|1)$ WZW model is interesting from the point of view of physics, because in some of the articles it has attracted considerable attention:
By studying maximally symmetric branes in the $GL(1|1)$ WZW model it
was shown that such branes are localized along (twisted) super-conjugacy classes \cite{creutzig1} (see also \cite{creutzig2}).
The correlators of the model through a
free field representation were constructed out in \cite{creutzig3}, then,
by investigating some properties of the theory it was shown that
some of the model correlators can be contained logarithmic singularities.
Generally, WZW models on Lie supergroups present themselves as an
ideal playground to extend many of the beautiful results of unitary rational conformal
field theory to logarithmic models. Even the simplest models are mathematically rich and
physically relevant.
In addition, the existence of super Poisson-Lie symmetry is the most
important feature of the $GL(1|1)$ WZW model \cite{ER7}.

We hope that in future it will be possible to
find other YB deformed WZW models, especially for physically interesting backgrounds.
As a future direction, it would be interesting to get the YB deformations of the WZW models on Lie supergroups in higher dimensions
such as the $OSP(1|2)$ and $OSP(2|2)$
by following our present analysis and method.
However, our results in the present work can still provide insights into (generalized) supergravity solutions.
For this purpose, one must generalize the generalized supergravity equations to supermanifolds.
Some of these problems are currently under investigation.

\subsection*{Acknowledgements}

This work has been supported by the research vice
chancellor of Azarbaijan Shahid Madani University under research fund No. 97/231.


\end{document}